\documentclass[a4paper]{jpconf}
\usepackage{graphicx}
\def\yohkoh{{\sl Yohkoh}}
\def\hinode{{\sl Hinode}}
\def\soho{{\sl SOHO}}
\def\sdo{{\sl SDO}}
\def\stereo{{\sl STEREO}}
\def\sdo{{\sl SDO}}
\def\iris{{\sl IRIS}}
\def\smei{{\sl SMEI}}
\def\psp{{\sl PSP}}

\def\kms{km~s$^{-1}$}

\newcommand{\ltsim}{\raisebox{-1.0ex}{$\stackrel{\textstyle<}{\sim}$}}
\def\heii{He~{\sc ii}}
\def\siiv{S~{\sc iv}}

\def\al{Alfv\'{e}n}

\begin{document}
\title{Possible Evolution of Minifilament-Eruption-Produced Solar Coronal Jets, Jetlets, and Spicules, 
into Magnetic-Twist-Wave ``Switchbacks" Observed by the Parker Solar Probe (\psp)}

\author{Alphonse C. Sterling$^1$, Ronald L. Moore$^{2, 1}$, Navdeep K. Panesar$^{3, 4}$, 
and Tanmoy Samanta$^{5, 6}$}

\address{$^1$NASA/Marshall Space Flight Center, Huntsville, AL, 35806, USA}
\address{$^2$Center for Space Plasma and Aeronomic Research (CSPAR), UAH, Huntsville, AL 35805, USA}
\address{$^3$Bay Area Environmental Research Institute, NASA Research Park, Moffett Field, CA 94035, USA}
\address{$^4$Lockheed Martin Solar and Astrophysics Laboratory, 3251 Hanover Street, Bldg. 252, Palo Alto, CA 94304, USA}
\address{$^5$Department of Physics and Astronomy, George Mason University, Fairfax, VA 22030, USA}
\address{$^6$Johns Hopkins University Applied Physics Laboratory, Laurel, MD 20742, USA)}

\ead{alphonse.sterling@nasa.gov}

\begin{abstract}

Many solar coronal jets result from erupting miniature-filament (``minifilament") magnetic  flux
ropes that reconnect with encountered surrounding far-reaching field.  Many of those minifilament flux ropes 
are apparently built and triggered to erupt  by magnetic flux  cancelation.  If that cancelation (or some other
process) results in the flux rope's field having twist, then the reconnection with the far-reaching field
transfers much of that twist to that reconnected far-reaching field.  In cases where that surrounding 
field is open, the twist
can propagate to far distances from the Sun as a magnetic-twist \al ic pulse.  We argue that such pulses
from jets could be the kinked-magnetic-field structures known as ``switchbacks," detected in the
solar wind during perihelion passages of the Parker Solar Probe (\psp\@).  For typical
coronal-jet-generated \al ic pulses, we expect that the switchbacks would flow past \psp\  with a
duration of several tens of minutes; larger coronal jets might produce switchbacks with passage
durations $\sim$1\,hr.  Smaller-scale jet-like features on the Sun known as ``jetlets" may be
small-scale versions of coronal jets, produced in a similar manner as the coronal jets.  We estimate
that switchbacks from jetlets would flow past \psp\ with a duration of a few minutes.  Chromospheric
spicules are jet-like features that are even smaller than jetlets.  If some portion of their population
are indeed very-small-scale versions of coronal jets, then we speculate that the same processes could
result in switchbacks that pass \psp\ with durations ranging from about $\sim$2\,min down to tens of seconds.

\end{abstract}

\section{Introduction}
\label{sec-introduction}

Solar coronal jets are transient features in the Sun's outer atmosphere, the corona, that were first  studied in
detail using observations from the X-ray telescope on the \yohkoh\ satellite (\cite{shibata.et92,shimojo.et96}),
which was launched in 1991.  They grow  to be long and narrow (typically around $50{,}000 \times 10{,}000$~km) over
a short period of time $\sim$10\,min, for those occurring in polar coronal holes
(\cite{cirtain.et07,savcheva.et07}).  They are also common in quiet Sun and active regions.  Since \yohkoh,  they
have been observed extensively in X-rays from the X-ray telescope (XRT)  on \hinode, and in various EUV filters
from the Atmospheric Imaging Assembly (AIA) on  the Solar Dynamics Observatory (\sdo) satellite.  They have also
been studied with other instruments and in  different wavelength ranges.  Their properties have been reviewed
elsewhere (\cite{shibata.et11,raouafi.et16,hinode.et19}). 

{\it Switchbacks} are localized structures in the solar wind where the magnetic-field direction 
undergoes large variation.  Although detected and discussed previously (e.g., 
\cite{kahler.et96,yamauchi.et04,suess07}), they have recently become of wider interest due to a
series of detailed {\it in situ} observations by the Parker Solar Probe (\psp) very close to the Sun,
during its perihelion passages of $\sim$35\,$R_{\odot}$ in 2018 November and 2019 April. The kinks in  the
magnetic field of the switchbacks are sometimes extreme enough for the field's radial component to reverse for  a period of
time.  They are possibly \al ic pulses that are propagating in the solar wind, whereby they pass \psp\ as
they move outward from the Sun riding on the solar wind.  Their durations vary widely, with \psp-crossing 
times ranging from less than one second to over an hour (\cite{dudok.et20}).

Here we consider how coronal jets and similar features might be the source of the \psp\ switchbacks, as 
mentioned previously 
(\cite{bale.et19,dudok.et20}), and discussed in more detail in \cite{horbury.et20}.  Much of the
presentation here is based on the investigation in \cite{sterling.et20b}.  Although not addressed
further here, we point out that there are other ideas under discussion for switchback production 
(e.g., \cite{dudok.et20}).

\section{From Minifilament Eruptions to Solar Coronal Jets}
\label{sec-jets}

Early models of jets assumed that they formed when new magnetic flux emerging from the photosphere
reconnected with encountered far-reaching coronal field (\cite{shibata.et92,yokoyama.et95}).  Later observations
however showed that jets, especially those in coronal holes and quiet regions, frequently 
originate from sites of flux convergence rather than emergence 
(e.g., \cite{huang.et12,shen.et12,young.et14a,adams.et14}), and that they frequently occur 
as a result of eruption of a miniature filament (``minifilament") (e.g., \cite{shen.et12,adams.et14}).
\cite{sterling.et15} showed that polar coronal hole jets commonly result from minifilament eruptions.
Subsequent work (e.g., \cite{panesar.et16a,panesar.et17,panesar.et18a,mcglasson.et19}) argues that flux cancelation 
builds the minifilament flux rope and triggers its eruption in most cases.  (Work of \cite{kumar.et19} confirms
that many coronal hole jets result from minifilament eruptions, but argues that the eruptions often 
result from flux convergence and/or shearing motions rather than cancelation.)   

A minifilament-eruption-model, schematically presented by \cite{sterling.et15}, describes how 
the minifilament flux-rope eruption can drive the jet
and concurrently explain a brightening that often appears at the base of the jet and off to one side of the
spire (Fig.\,1).  They argue that the jet is a small-scale version of the better-known larger filament eruptions
that often produce coronal mass ejections (CMEs), and that the base-edge brightening is a 
small-scale version a typical solar-flare brightening (i.e.\ the brightening from the flare arcade that forms 
in the wake of larger-scale filament eruptions that sometimes produce CMEs).  Numerical modeling confirms 
the plausibility of this schematic (\cite{wyper.et17,wyper.et18a}).

In active region coronal jets, minifilament eruptions are sometimes not as obvious as in many jets in  quieter solar
regions.  Nonetheless, flux cancelation leading to minifilament flux-rope eruptions that  make jets appears to be the basic
process driving many active region jets also (e.g.,
\cite{sterling.et16b,sterling.et17,joshi_schmieder.et17,shen.et17,hong.et17,bucik.et18,miao.et19,solanki.et20}). 
The lack of obvious erupting minifilaments (and some other features) in some active region jets might be a consequence of the
complex magnetic field arrangement and rapid evolution in those regions  compared to quieter regions \cite{sterling.et17}.

Jets also frequently show twisting motions (e.g,
\cite{pike.et98,harrison.et01,patsourakos.et08,kamio.et10,moore.et10,moore.et13,schmieder.et13,cheung.et15,panesar.et16a,joshi.et18,liu.et19}).
The minifilament eruption model (Fig.\,1) is consistent with this if twist first builds up in the minifilament flux rope prior
to its eruption, and then some of that twist is transferred onto the open field via magnetic reconnection
(\cite{shibata.et86,moore.et15,yang.et19}).

\begin{figure} 
\hspace*{0.0cm}\includegraphics[angle=0,scale=0.95]{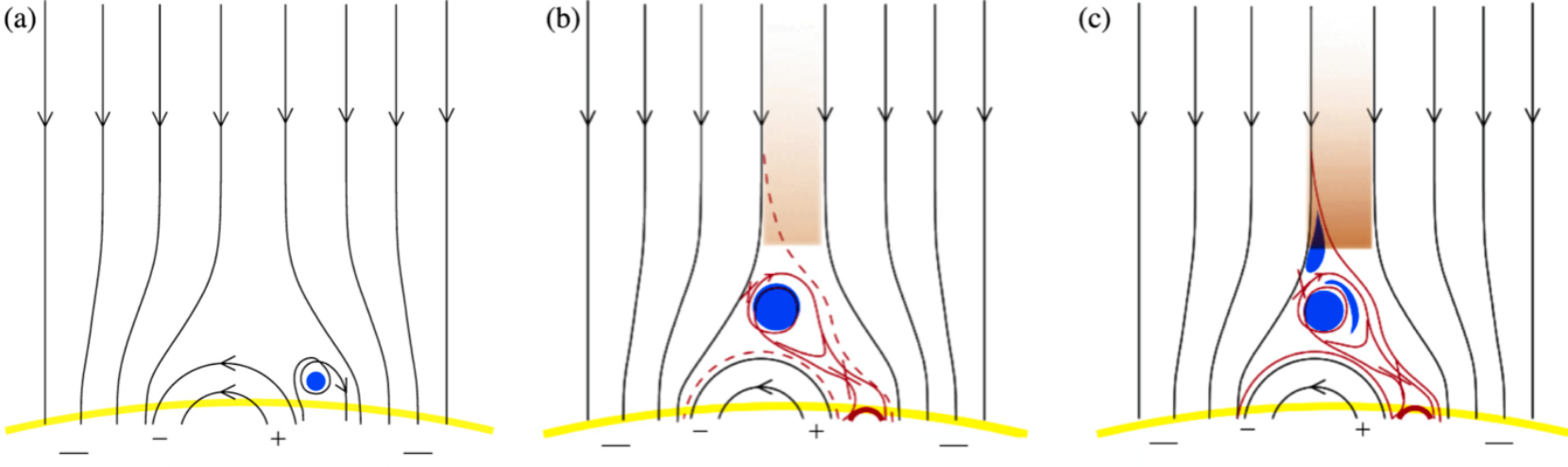}
\caption{Schematic showing the {\it minifilament eruption model for coronal jets}; originally  presented in
\cite{sterling.et15}, this version is from \cite{sterling.et20b}.  This represents a  near-limb 2D
cross-sectional view of the jetting region, with the yellow curve representing the solar limb.  Black lines
represent magnetic field lines prior to magnetic reconnection, and red lines represent reconnected field 
lines.  In the most general jet model the vertical lines can represent either open or
far-reaching closed field, but here we only consider the open-field case since we are assuming that the jet
effects extend into the distant heliosphere. (a) A minifilament  (blue disk) is in a magnetic bipole, 
pictured here as residing on the right-hand side of  a
larger bipole (see the plus- and minus-polarities at the bottom of the panel).  The curled field line
encircling the minifilament indicates non-potential twist in the field holding the  minifilament material.  (b) The
minifilament field erupts, running into and undergoing interchange reconnection with opposite-polarity open 
field, leading to new field lines (dotted) including new open field that guides a hot jet (shaded region), 
and with field below the minifilament undergoing reconnection (``internal reconnection''
\cite{sterling.et15}) and forming a frequently observed brightening on one side of the jet's 
base (thick red semicircle).  (c) As the minifilament eruption
progresses, the interchange reconnection can eat into the minifilament flux rope enough to release cool
minifilament material to travel outward along the jet spire.  That interchange reconnection also 
transfers twist that is in the erupting minifilament flux rope to the open field, resulting in a
frequently observed spinning jet spire.  See
\cite{sterling.et15,sterling.et16b,sterling.et17,moore.et18} for more details.  This picture has been modeled
numerically (\cite{wyper.et17,wyper.et18a}).} 
\end{figure} 


\section{From Solar Coronal Jets to \psp\ Switchbacks}
\label{sec-switchbacks}

From the above discussion, we can say that: Coronal jets 
are frequently produced by minifilament flux-rope eruptions, the minifilament flux ropes
often appear to be made by magnetic flux cancelation, and the eruptions apparently 
are at least many times triggered by flux
cancelation.  It is plausible that canceling sheared fields form
minifilament flux ropes containing twist, and that when the minifilament flux rope erupts and makes 
the jet, it
transfers some of that twist to the open coronal field via magnetic reconnection.  This puts an \al ic 
twist wave onto the jet-spire field, resulting in the frequently observed jet-spire rotation.

A plausible interpretation of switchbacks observed by \psp\ is that they are propagating \al ic twist-wave
packets riding on the solar wind.  Here we consider whether the just-described twists of the jets 
could become switchbacks.

First we consider evidence that effects of coronal jets extend beyond the inner corona and into 
the outer corona and inner heliosphere.  

Several studies (e.g., 
\cite{wang.et98,nistico.et09,paraschiv.et10,hong.et11,hong.et13,moore.et15,sterling.et16b})
show that coronagraph-observed ``narrow CMEs" (of width from Sun-disk center of $\ltsim$5$^\circ$), 
also called ``white-light jets," originate from coronal jets.
Similar features extending from coronal jets have been observed in eclipse images (\cite{hanaoka.et18,alzate.et17b}).  
Coronal jets (or coronal-jet remnants) have been detected well into the heliosphere (a large fraction of an AU)
via 3D reconstructions using Solar Mass Ejection Imager (\smei) observations (\cite{yu.et14,yu.et16}).
They have also been detected in interplanetary space from the Hi1 Heliospheric Imager on 
the \stereo\ spacecraft \cite{sterling.et20b}.  This is strong evidence that the influence 
of at least some coronal jets can persist out to the distances of \psp\ perihelia, and beyond.

A natural follow-up question is: Which particular jets are most likely to persist out
to \psp\ locations?  While we cannot yet supply an exact answer to this question,
work by \cite{moore.et15} found that the coronal jets that showed up as white-light
jets in \soho/LASCO C2 coronagraph images were those that tended to display comparatively larger twisting
motions in \sdo/AIA 304~\AA\ \heii\ images.  More specifically, in earlier studies of random 
polar jets, \cite{moore.et13} found that $\sim$80\% of the 29 polar coronal hole jets they studied
displayed spinning amounts of $\ltsim$0.5 turns; in contrast, the 14 polar coronal hole jets known
to be counterparts of white-light jets all had twist amounts of between 0.5 and 2.5 turns.  Thus,
this suggests that the jets with relatively large twist are most likely to make detectable 
white-light jets.  Moreover, \cite{moore.et15} found evidence that the twists of the coronal jets
persisted out into the LASCO C2 corona as swaying motions in the corresponding white-light jets.

Under the assumption then that a minifilament flux rope that contains twist erupts to produce a jet on reconnected
coronal field that extends out into the heliosphere, we can expect that much of the  twist of the
minifilament field is transferred onto the open field. This twist may then  continue to propagate
outward, driving a density enhancement, and appearing as a white-light jet and a density 
enhancement in  interplanetary space.  (Our basic
argument here is independent of whether the erupting minifilament field contains cool minifilament
material. It might be that the addition of cool, denser, minifilament material is required in order for the
feature to appear as a white-light jet, but this is not yet established.)

The twist that has been transferred to the open field should then propagate outward as a magnetic-twist pulse 
at the \al\ speed.  In the inner heliosphere, the \al\ speed (=$B/(4\pi \rho)^{0.5}$, where $B$ is
the field strength and $\rho$ the plasma mass density) decreases with distance from the Sun.  In the
corona we can expect an \al\ speed of 1000~\kms\ (e.g.\ in coronal holes), while \psp\
found an \al\ speed of $\sim$100~\kms\ at its perihelion of $\sim$35\,$R_{\odot}$ (\cite{bale.et19}),
confirming a substantial gradient in the \al\ speed with distance from the Sun.

The effect of this gradient is that an \al-wave pulse of finite length will feel different local \al\ speeds along
its extent.  An \al\ wave packet propagates at different speeds
between its ends: faster nearer the Sun and slower farther from the Sun.  The result is a contraction, or
``steepening,"  of the pulse packet wave with distance (and hence time) as it moves out from the Sun (Fig.\,2).

Following \cite{sterling.et20b}, taking a coronal \al\ speed of 1000\,\kms\ and a typical jet lifetime
of 10\,min, we can expect the time for the minifilament flux rope to transfer its twist to the open coronal field
to be $\sim$600\,s, so that the length of the \al\ twist-wave pulse in the corona would be $\sim$600{,}000\,km.
During the first \psp\ perihelion, the solar wind speed at \psp\ was $\sim$300\,\kms\ (\cite{kasper.et19}).
With the jet-produced \al\ pulse riding on the wind at the local \al\ speed of 100\,\kms, the pulse
will pass \psp\ at $\sim$400\,\kms.  So if there were negligible contraction of the pulse between the
corona and \psp\, we would expect the time for the passage to be about 25\,min; with contraction, it is 
less than this.

\begin{figure}
\hspace*{3.3cm}\includegraphics[angle=0,scale=0.5]{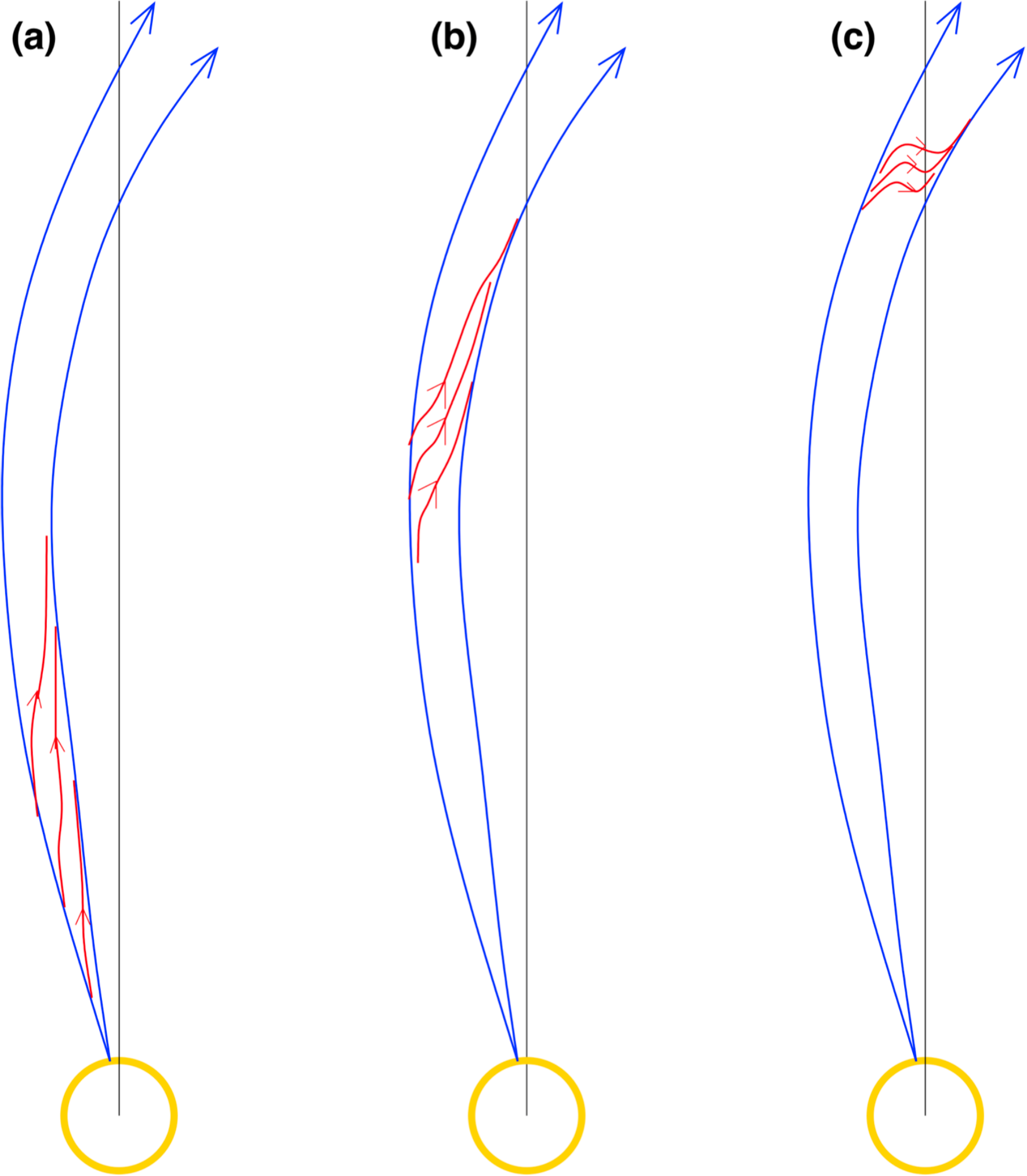}
\caption{This figure is from \cite{sterling.et20b}.  Schematic showing a Parker spiral heliospheric
field (blue lines) that are curved relative to a radial line (black) extending out from the Sun 
(yellow circles).  This is an extension of Fig.\,1, 
where the twist transferred to the reconnected open coronal field through the interchange reconnection in Figs.\,1(b) 
and~1(c) is the twist that is shown here in
the open field, and which propagates outward as an \al ic pulse.  
In (a),
the twist transferred to the open field becomes the red disturbance, that appears as a
low-pitch twist wave packet moving outward (when launched from a coronal jet the radial extent of 
the twist packet is estimated be comparable 
to a solar radius, and so 
its extent is exaggerated
by a factor of a few times compared to the Sun  in this schematic representation).  Panel~(b) shows 
how the pitch of the disturbance could increase, as described in the text, as it moves further 
from the Sun and into a regime of lower \al\ speed compared 
to that lower in the corona.  In (c), this pitch-angle steepening of the impulse continues as it moves even 
further from the Sun.  We suggest that this \al ic disturbance could appear as a 
``switchback" by the time it encounters \psp\@.}
\end{figure}

\section{A Possible Source for Shorter-Duration Switchbacks: Smaller-scale ``Coronal Jets."}
\label{sec-smalljets}

From the above arguments, and given the variety of jet-lifetime durations (100---2000\,s according 
to \cite{kim.et07}), and with the expected contraction of the \al\ pulse with travel time, it is
plausible that coronal jets from minifilament eruptions could explain switchbacks of durations lasting a 
few minutes to a few tens of minutes.  Given arguments that many larger-scale eruptions occur in
circumstances similar to coronal jets (e.g., \cite{joshi.et17a,sterling.et18}), we can expect that there
will also be some longer-duration switchbacks ejected to the \psp\ perihelion location.  As mentioned
in \S\ref{sec-introduction}, however, switchbacks durations range over less than one second to over
an hour; if typical coronal jets account for switchbacks 
of a few minutes or tens of 
minutes, then somewhat-larger-scale coronal jets (or small-scale solar eruptions) might 
account for switchbacks of durations $\sim$1\,hr.
Could
the same scenario explain shorter-duration switchbacks?  It turns out that there is some evidence
that frequently occurring jets that are smaller-scale versions of typical coronal jets do occur on the 
Sun.  We now briefly discuss those in the context of switchbacks.

\subsection{A Possible Jetlet-Switchback Connection}
\label{subsec-jetlets}

``Jetlets" are jet-like features that occur on a smaller scale than coronal jets.  They were identified 
(\cite{raouafi.et14}) in AIA EUV images at the base of coronal plumes, occurring at sites where magnetic  minority
polarity elements cancel with surrounding field.  That \cite{raouafi.et14}  study argued that the jetlets and other cancelation-induced
transients are the main energy source for the plumes.  Subsequently, \cite{panesar.et18b}  examined ten jetlets, using both 
EUV images from AIA and UV images from the Interface Region Imaging Spectrograph
(\iris) spacecraft.  They found that the jetlets occurred at the boundaries of the magnetic network, and
that they were not restricted to plume locations.  According to \cite{panesar.et18b}, jetlets have on average widths of about
3000\,km and live about 3\,min, values that are about three-times smaller than the corresponding values
for typical coronal jets (\cite{savcheva.et07}). Jetlet speeds of $\sim$70\,\kms\ (\cite{panesar.et18b}) are comparable to those 
found for coronal jets 
($\sim$70---100\,\kms, \cite{panesar.et16b,panesar.et18a}).  \cite{panesar.et18b} found them to have average
maximum spire lengths of 27{,}000\,km and 16{,}000\,km when measured respectively in EUV (AIA 171\,\AA) and UV
(\iris\ \siiv\ slit-jaw images); which is substantially shorter than typical spire lengths of coronal jets observed in
X-rays (50{,}000\,km, \cite{savcheva.et07}).

\cite{panesar.et18b} and \cite{panesar.et19} argue that many jetlets are small-scale version of  coronal
jets.  Although no minifilament eruption has been observed at their base, they do share  other
similarities with jets, including evidence that they occur at sites of magnetic flux cancelation, and
that brightenings are frequently observed at the base of the jetlets near the time of their onset.
There are also hints that some jetlet spires may show spinning motions as they grow.

Building on this evidence that jetlets are small-scale coronal jets, if we further assume that  they are
produced as described in \S\ref{sec-jets}, that is, cancelation of sheared magnetic  flux builds a
twisted flux rope that erupts and produces the jetlet (Fig.\,1) and imparts spin to it, then we can speculate that
the jetlets might also be a source for switchbacks.  (N. Raouafi (2019, private communication) speculated
that jetlets are the source of switchbacks; also see \cite{bale.et19}.)  Following
the  discussion in \S\ref{sec-switchbacks}, we can estimate the maximum size and \psp-passage duration 
of the jetlet switchbacks.  With a 3-min lifetime and coronal \al\ speed of 1000\,\kms, a  $\sim$180{,}000\,km
\al ic pulse packet would be expected to be loaded onto the low-coronal open field at the site of the jetlet's
formation.  A pulse of this size will pass \psp\ in 7.5\,min, even without any contraction; with contraction  we can expect the
durations of the jetlet-generated switchbacks to be of order a few minutes.

\subsection{A Possible Spicule-Switchback Connection}
\label{subsec-spicules}

Could switchbacks be made the same way on even smaller-size scales than jetlets?  Extending the same concept of
jetlets being coronal jets occurring on reduced size scales, it could be that at least some chromospheric spicules are 
made in a similar manner as jets but on a size scale even smaller than jetlets, and hence produce 
switchbacks of even shorter \psp-passage durations.

Spicules have been observed for a long time with a wide variety of instruments.  They are jet-like 
features, but even delineating  their properties has been a challenge, as those observed properties
are dependent on both the instruments and wavelengths used, also on the resolution and time 
cadence of any particular instrument. Some general approximate numbers are that they have maximum 
lengths of a few thousand kms, widths of a few
hundred km, and lifetimes of a few minutes, as discussed in several reviews and summary papers 
(e.g., \cite{beckers68,sterling00,depontieu.et07a,tsiropoula.et12,hinode.et19,samanta.et19}).  
There are many ideas for how they are created, as discussed in the same review papers, while some
more recent investigations include \cite{iijima.et17} and \cite{martinezsykora.et17}.

It has been suggested (\cite{sterling.et16a}) that at least some spicules might be small-scale versions 
of coronal jets.    Recent high-resolution on-disk spicule observations 
(\cite{samanta.et19,sterling.et20a}) furthered this argument.  It has long been suspected 
(e.g., \cite{pasachoff.et68}) that some spicules
spin, and now this has been shown for some spicules spectroscopically  (\cite{depontieu.et12}).  
Also, recent high-resolution
observations by \cite{samanta.et19} show that magnetic activity, at least some of which might be flux 
cancelation, occurs at the base of at least some spicules.  These two aspects - spin and  possible flux
cancelation - along with their general jet-like appearance, are qualities of spicules that are similar
to many coronal jets.  ``{\it Micro}filament" eruptions making spicules, potentially corresponding 
to the minifilament eruptions that often make coronal jets, have not been compellingly 
detected, and therefore the idea that
spicules are scaled-down coronal jets is still speculative.  Nonetheless, a simple extrapolation
of number of erupting filament-like features on the Sun as a function of the size of those 
erupting filaments (\cite{sterling.et16a}) suggests that at least some coronal-jet-like features of the
spicule-size category could be present in the chromosphere, independent of the question of 
whether that mechanism would make up a majority of spicules or only a small fraction of the
total spicule population.

If some spicule-like features are produced by microfilament eruptions by the same process that
makes many coronal jets (Fig.\,1), then we can ask what a corresponding switchback might look
like from that population of spicules.  A spicule's microfilament eruption timescale might be
30\,s---1\,min, based on the candidate erupting microfilaments identified in \cite{sterling.et20a}. 
This gives a twist-pulse length in the corona of $\sim$30{,}000---60{,}000\,km.
Sticking with our earlier assumptions and just considering
a pulse of this size, it would pass \psp\
in just 75---150\,s, and so $\sim$2\,min.  With contraction as envisaged in Fig.\,2, this time 
period - at least for the part of the pulse displaying the most rapid magnetic-field-direction 
changes - would be $\sim$1\,min, down to perhaps as short as a few tens of seconds.

It has already been proposed that a reconnection-like process simultaneously produces spicules 
and generates propagating coronal waves within plumes (\cite{samanta.et15,pant.et15}), which 
could be consistent with our view in Fig.\,2. 
In addition, high spatial- and temporal-resolution studies show that incompressible transverse motions 
occur ubiquitously in mottles/spicules (\cite{depontieu.et07c,jess.et09,kuridze.et12,kuridze.et13,morton.et13,morton14,zaqarashvili.et09});
whether these motions result from dynamics such as low-altitude reconnections and twist transfer
as envisioned in Fig.\,1 is a topic for future investigation.

\section{Discussion and Conclusions}
\label{sec-conclusions}

There is now excellent evidence that many quiet Sun and coronal hole jets are made by 
minifilament eruptions.  There is good evidence that in many cases the minifilament flux rope is built 
and triggered to erupt by magnetic flux cancelation.  In many cases, active region jets develop
and form the same way, but the process can be complicated by the more-complex and rapidly evolving
magnetic environment of active regions compared to quieter solar regions.  If the so-formed minifilament 
flux rope has twist prior to its eruption, then via the scenario of Fig.\,1, it will transfer much of that 
twist to the open field with which it reconnects, inducing the often-observed twisting motion
of the jet spire.  We have argued that the twist plausibly could propagate out to \psp-perihelia
locations as an \al ic twist-wave pulse.  \al-speed decrease in the heliosphere with radial distance from the Sun
could then lead to a steepening of the kink in the magnetic field in the pulse as it 
travels away from the Sun along a Parker spiral, as depicted in Fig.\,2.  For the case of coronal 
jets, we have argued that \psp\ could see the passage of that kinked \al ic pulse as a switchback
of likely duration of several tens of minutes.  Larger jets, and larger (but still small-scale) filament 
eruptions, might result in switchbacks with passage times of $\sim$1\,hr.

If jetlets work in the same way as these coronal jets, then they could similarly produce 
\psp-observed switchbacks of likely duration of the order of several minutes.  If we further speculate
that some portion of the spicule population are similarly produced scaled-down jets, then they
could result in \psp-observed switchbacks of likely duration of a couple of minutes, down to a few
tens of seconds.

Our work here does not directly explain very short duration switchbacks, down to a few seconds or
even less than a second.  It should be expected however that \psp\ sometimes skirts by (nicks the edge 
of) a jet-induced switchback
 (rather than encountering the full wave packet along the packet's direction of travel), substantially shortening 
the encounter duration.  In this sense, the durations that we have derived above can be viewed as upper limits
to the expected duration of the switchbacks encountered by \psp\@.

At this point in time, as the scale size decreases our story becomes more speculative.  For example, 
we have not unquestionably detected the equivalent of a cool-material minifilament eruption in jetlets 
or spicules.  This however could be a natural limitation of current observational capabilities.  Future 
investigations with current and future instruments should address further the question of the origin of these
features.

Also, future numerical simulations should address how \al ic twist-wave pulses launched in the low solar atmosphere will
evolve as they propagate out to \psp\ locations.  Numerical simulations by \cite{tenerani.et20} show that \al ic
impulses can indeed maintain their integrity for estimated distances of tens of $R_{\odot}$ in the solar wind, given calm-enough
solar-wind conditions.  Codes such as that of \cite{tenerani.et20}, and others, for example codes those of
\cite{lionello.et16} and \cite{roberts.et18} that model magnetic connections between the photosphere and the
heliosphere, might be appropriate for investigations into the ideas discussed here, provided such codes include modeling
of the minifilament-field eruption that drives the jet.

Our schematic in Fig.\,2 explains how the twist pulse might steepen into a $\sim$90-degree swing in field direction 
from the solar radial direction (it might be slightly larger than 90 degrees from radial, due to the Parker-spiral angle). 
It does not however picture
an ``extreme" switchback, of swing angle substantially greater than 90 degrees.  Such large-angle switchbacks are relatively 
uncommon, but seemingly more frequent further from the Sun (\cite{mozer.et20}).  It is possible that the twist wave
that we envision might evolve into such a
state on small-enough length scales, where the wave pulse encounters a local solar wind inhomogeneity, for example. 
Numerical simulations incorporating appropriate solar-wind physics might be able to address this question.
On the other hand, recent \psp\ investigations show that the typical switchback field-rotational angle increases with
radial distance from the Sun (\cite{mozer.et20}), which is consistent with the picture of Fig.\,2.

If coronal jets and similar jet-like features are the source of the switchbacks, then we would expect the rate of
switchbacks encountered by \psp\ with closer and closer perihelia continually to remain high or increase, 
while the average swing angle continually decreases.   If other ideas for
switchback origin predict different trends of switchback frequency and swing angle with distance from 
the Sun  (for example, some ideas might predict the switchbacks to start at a certain radial distance from the Sun,
while our idea predicts their existence from the low-corona outward), then the findings
from upcoming \psp\ perihelia might be used to help narrow candidate ideas for switchback generation.

\ack
This work was supported by funding from the NASA Heliophysics Guest Investigators program, and 
the MSFC Hinode project.

\section*{References}

\bibliography{icns_proceedings_2020_ah}

\providecommand{\newblock}{}
\begin{thebibliography}{10}
\expandafter\ifx\csname url\endcsname\relax
  \def\url#1{{\tt #1}}\fi
\expandafter\ifx\csname urlprefix\endcsname\relax\def\urlprefix{URL }\fi
\providecommand{\eprint}[2][]{\url{#2}}

\bibitem{shibata.et92}
Shibata K, Ishido Y, Acton L~W, Strong K~T, Hirayama T, Uchida Y, McAllister
  A~H, Matsumoto R, Tsuneta S, Shimizu T, Hara H, Sakurai T, Ichimoto K,
  Nishino Y and Ogawara Y 1992 {\em Publications of the Astronomical Society of
  Japan\/} {\bf 44} L173

\bibitem{shimojo.et96}
Shimojo M, Hashimoto S, Shibata K, Hirayama T, Hudson H~S and Acton L~W 1996
  {\em Publications of the Astronomical Society of Japan\/} {\bf 48} 123

\bibitem{cirtain.et07}
Cirtain J~W, Golub L, Lundquist L, van Ballegooijen A, Savcheva A, Shimojo M,
  DeLuca E, Tsuneta S, Sakao T, Reeves K, Weber M, Kano R, Narukage N and
  Shibasaki K 2007 {\em Science\/} {\bf 318} 1580

\bibitem{savcheva.et07}
Savcheva A, Cirtain J, Deluca E~E, Lundquist L~L, Golub L, Weber M, Shimojo M,
  Shibasaki K, Sakao T, Narukage N, Tsuneta S and Kano R 2007 {\em Publications
  of the Astronomical Society of Japan\/} {\bf 59} 771

\bibitem{shibata.et11}
Shibata K and Magara T 2011 {\em LRSP\/} {\bf 8} 6

\bibitem{raouafi.et16}
Raouafi N~E, Patsourakos S, Pariat E, Young P~R, Sterling A~C, Savcheva A,
  Shimojo M, Moreno-Insertis F, DeVore C~R, Archontis V, Török T, Mason H,
  Curdt W, Meyer K, Dalmasse K and Matsui Y 2016 {\em Space Science Reviews\/}
  {\bf 201} 1

\bibitem{hinode.et19}
{Hinode Review Team}, {Khalid} A~J, {Patrick} A, {Baker} D, R B~L and {et al}
  2019 {\em Publications of the Astronomical Society of Japan\/} {\bf 71} id.R1

\bibitem{kahler.et96}
Kahler S~W, Crooker N~U and Gosling J~T 1996 {\em Journal of Geophysical
  Research\/} {\bf 101} 24373

\bibitem{yamauchi.et04}
Yamauchi Y, Suess S~T, Steinberg J~T and Sakurai T 2004 {\em Journal of
  Geophysical Research: Space Physics\/} {\bf 109} A03104

\bibitem{suess07}
Suess S 2007 Status of knowledge after ulysses and soho {\em Proceedings of The
  Second Solar Orbiter Workshop\/} ed Marsch E;~Tsinganos K and Marsden
  R;~Conroy L (Noordwijk, Netherlands: European Space Agency) p 641

\bibitem{dudok.et20}
Dudok~de Wit T, Krasnoselskikh V~V, Bale S~D, Bonnell J~W, Bowen T~A, Chen
  C~H~K, Froment C, Goetz K, Harvey P~R, Jagarlamudi V~K, Larosa A, MacDowall
  R~J, Malaspina D~M, Matthaeus W~H, Pulupa M, Velli M and Whittlesey P~L 2020
  {\em Astrophysical Journal Supplement Series\/} {\bf 246} 39

\bibitem{bale.et19}
Bale S~D, Badman S~T, Bonnell J~W, Bowen T~A, Burgess D, Case A~W, Cattell C~A,
  Chandran B~D~G, Chaston C~C, Chen C~H~K, Drake J~F, Dudok~de Witt T, Eastwood
  P, Ergun R~E, Farrell W~M, Fong C, Goetz K, Goldstein M, Goodrich K~A, Harvey
  P~R, Horbury T~S, Howes G~G, Kasper J~C, Kellogg P~J, Klimcuk J~A, Korreck
  K~E, Krasnoselskikh V~V, Krucker S, Laker R, Larson D~E, MacDowall R~J,
  Maksimovic M, Malaspina D~M, Martinez-Oliveros J, McComas D~J, Meyer-Vernet
  N, Moncuquet M, Mozer F~S, Phan T~D, Pulupa M, Raouafi N~E, Salem C, Stansby
  D, Stevens M, Szabo A, Velli M, Woolley T and Wygant J~R 2019 {\em Nature\/}
  {\bf 576} 237

\bibitem{horbury.et20}
Horbury T~S, Woolley T, Laker R, Matteini L, Eastwood J, Bale S~D, Velli M,
  Chandran B~D~G, Phan T, Raouafi N~E, Goetz K, Harvey P~R, Pulupa M, Klein
  K~G, Dudok~de Wit T, Kasper J~C, Korreck K~E, Case A~W, Stevens M~L,
  Whittlesey P, Larson D, MacDowall R~J, Malaspina D~M and Livi R 2020 {\em
  Astrophysical Journal Supplement Series\/} {\bf 246} 45

\bibitem{sterling.et20b}
Sterling A~C and Moore R~L 2020 {\em Astrophysical Journal Letters, {\it in
  press}\/}

\bibitem{yokoyama.et95}
Yokoyama T and Shibata K 1995 {\em Nature\/} {\bf 375} 42

\bibitem{huang.et12}
Huang Z M, S M, Doyle J~G and Lamb D~A 2012 {\em Astronomy and Astrophysics\/}
  {\bf 548} 62

\bibitem{shen.et12}
Shen Y, Liu Y, Su J and Deng Y 2012 {\em Astrophysical Journal\/} {\bf 745} 164

\bibitem{young.et14a}
Young P~R and Muglach K 2014 {\em Solar Physics\/} {\bf 289} 3313

\bibitem{adams.et14}
Adams M, Sterling A~C, Moore R~L and Gary G~A 2014 {\em Astrophysical
  Journal\/} {\bf 783} 11

\bibitem{sterling.et15}
Sterling A~C, Moore R~L, Falconer D~A and Adams M 2015 {\em Nature\/} {\bf 523}
  437

\bibitem{panesar.et16a}
Panesar N~K, Sterling A~C, Moore R~L and Chakrapani P 2016 {\em Astrophysical
  Journal\/} {\bf 832L} 7

\bibitem{panesar.et17}
Panesar N~K, Sterling A~C and Moore R~L 2017 {\em Astrophysical Journal\/} {\bf
  844} 131

\bibitem{panesar.et18a}
Panesar N~K, Sterling A~C and Moore R~L 2018 {\em Astrophysical Journal\/} {\bf
  853} 189

\bibitem{mcglasson.et19}
McGlasson R~A, Panesar N~K, Sterling A~C and Moore R~L 2019 {\em Astrophysical
  Journal\/} {\bf 882} 16

\bibitem{kumar.et19}
Kumar P, Karpen J~T, Antiochos S~K, Wyper P~F, DeVore C~R and DeForest C~E 2018
  {\em Astrophysical Journal\/} {\bf 873} 93

\bibitem{wyper.et17}
Wyper P~F, Antiochos S~K and DeVore C~R 2017 {\em Nature\/} {\bf 544} 452

\bibitem{wyper.et18a}
Wyper P~F, DeVore C~R and Antiochos S~K 2018 {\em Astrophysical Journal\/} {\bf
  852} 98

\bibitem{sterling.et16b}
Sterling A~C, Moore R~L, Falconer D~A, Panesar N~K, Akiyama S, Yashiro S and
  Gopalswamy N 2016 {\em Astrophysical Journal\/} {\bf 821} 100

\bibitem{sterling.et17}
Sterling A~C, Moore R~L, Falconer D~A, Panesar N~K and Martinez F 2017 {\em
  Astrophysical Journal\/} {\bf 844} 28

\bibitem{joshi_schmieder.et17}
Joshi R, Schmieder B, Chandra R, Aulanier G, Zuccarello F~P and Uddin W 2017
  {\em Solar Physics\/} {\bf 292} 152

\bibitem{shen.et17}
Shen Y, Liu Y~D, Su J, Qu Z and Tian Z 2017 {\em Astrophysical Journal\/} {\bf
  851} 67

\bibitem{hong.et17}
Hong J, Jiang Y, Yang Y, Li H and Xu Z 2017 {\em Astrophysical Journal\/} {\bf
  835} 35

\bibitem{bucik.et18}
Bu{\u c}{\' i}k R, Innes D~E, Mason G~M, Wiedenbeck M~E, G{\' o}mez-Herrero R
  and Nitta N~V 2018 {\em Astrophysical Journal\/} {\bf 852} 76

\bibitem{miao.et19}
Miao Y, Liu Y, Shen Y~D, Elmhamdi A, Kordi A~S, Li H~B, Abidin Z~Z and Tian Z~J
  2019 {\em Astrophysical Journal\/} {\bf 877} 61

\bibitem{solanki.et20}
Solanki R, Srivastava A~K and Dwivedi B~N 2020 {\em Solar Physics\/} {\bf 295}
  27

\bibitem{pike.et98}
Pike C~D and Mason H~E 1998 {\em Solar Physics\/} {\bf 182} 333

\bibitem{harrison.et01}
Harrison R~A, Bryans P and Bingham R 2001 {\em Astronomy \& Astrophysics\/}
  {\bf 379} 324

\bibitem{patsourakos.et08}
Patsourakos S, Pariat E, Vourlidas A, Antiochos S~K and Wuelser J~P 2008 {\em
  Astrophysical Journal\/} {\bf 680} L73

\bibitem{kamio.et10}
Kamio S, Curdt W, Teriaca L, Inhester B and Solanki S~K 2010 {\em Astronomy and
  Astrophysics\/} {\bf 510} 1

\bibitem{moore.et10}
Moore R~L, Cirtain J~W, Sterling A~C and Falconer D~A 2010 {\em Astrophysical
  Journal\/} {\bf 720} 757

\bibitem{moore.et13}
Moore R~L, Sterling A~C, Falconer D~A and Robe D 2013 {\em Astrophysical
  Journal\/} {\bf 769} 134

\bibitem{schmieder.et13}
Schmieder B, Guo Y, Moreno-Insertis F, Aulanier G, Yelles~Chaouche L, Nishizuka
  N, Harra L~K, Thalmann J~K, Vargas~Dominguez S and Liu Y 2013 {\em Astronomy
  and Astrophysics\/} {\bf 559} A1

\bibitem{cheung.et15}
Cheung M~C~M, De~Pontieu B, Tarbell T~D, Fu Y, Tian H, Testa P, Reeves K~K,
  Mart{\'i}nez-Sykora J, Boerner P, W{\"u}lser J~P, Lemen J, Title A~M,
  Hurlburt N, Kleint L, Kankelborg C, Jaeggli S, Golub L, McKillop S, Saar S,
  Carlsson M and Hansteen V 2015 {\em The Astrophysical Journal\/} {\bf 801} 83

\bibitem{joshi.et18}
Joshi N~C, Nishizuka N, Filippov B, Magara T and Tlatov A~G 2018 {\em MNRAS\/}
  {\bf 476} 1286

\bibitem{liu.et19}
Liu J, Wang Y and Erd{\'e}lyi R 2019 {\em Frontiers in Astronomy and Space
  Sciences\/} {\bf 6} 44L

\bibitem{shibata.et86}
Shibata K and Uchida Y 1986 {\em Solar Physics\/} {\bf 178} 379

\bibitem{moore.et15}
Moore R~L, Sterling R~L and Falconer D~A 2015 {\em Astrophysical Journal\/}
  {\bf 806} 11

\bibitem{yang.et19}
Yang L, Yan X, Xue Z, Li T, Wang J, Li Q and Cheng X 2019 {\em Astrophysical
  Journal\/} {\bf 887} 239

\bibitem{moore.et18}
Moore R~L, Sterling R~L and Panesar N~K 2018 {\em Astrophysical Journal\/} {\bf
  859} 3

\bibitem{wang.et98}
Wang Y~M, Sheeley N~R, Jr, Socker, G D, Howard R~A, Brueckner G~E, Michels D~J,
  Moses D, Cyr S, C O, Llebaria A and Delaboudinière J~p 1998 {\em
  Astrophysical Journal\/} {\bf 508} 899

\bibitem{nistico.et09}
Nistic{\`o} G, Bothmer V, Patsourakos S and Zimbardo G 2009 {\em Solar
  Physics\/} {\bf 259} 87

\bibitem{paraschiv.et10}
Paraschiv A~R, Lacatus D~A, Badescu T, Lupu M~G, Simon S~Sandu S~G, Mierla M
  and Rusu M~V 2010 {\em Solar Physics\/} {\bf 264} 365

\bibitem{hong.et11}
Hong J, Jiang Y, Zheng R, Yang J, Bi Y and Yang B 2011 {\em Astrophysical
  Journal\/} {\bf 738L} 20

\bibitem{hong.et13}
Hong J, Jiang Y, Yang J, Zheng R, Bi Y, Li H, Yang B and Yang D 2013 {\em
  Research in Astronomy and Astrophysics\/} {\bf 13} 253

\bibitem{hanaoka.et18}
Hanaoka Y, Hasuo R, Hirose T, Ikeda A~C, Ishibashi T, Manago N, Masuda Y,
  Morita S, Nakazawa J, Ohgoe O, Sakai Y, Sasaki K, Takahashi K and Toi T 2018
  {\em Astrophysical Journal\/} {\bf 860} 142

\bibitem{alzate.et17b}
Alzate N, Habbal S~R, Druckm{\"u}ller M, Emmanouilidis C and Morgan H 2017 {\em
  Astrophysical Journal\/} {\bf 848} 84

\bibitem{yu.et14}
Yu H~S, Jackson B~V, Buffington A, Hick P~P, Shimojo M and Sako N 2014 {\em
  Astrophysical Journal\/} {\bf 784} 166

\bibitem{yu.et16}
Yu H~S, Jackson B~V, Yang Y~H, Chen N~H, Buffington A and Hick P~P 2016 {\em
  Journal of Geophysical Research: Space Physics\/} {\bf 121} 4985

\bibitem{kasper.et19}
Kasper J~C, Bale S~D, Belcher J~W, Berthomier M, Case A~W, Chandran B~D~G,
  Curtis D~W, Gallagher D, Gary S~P, Golub L, Halekas J~S, Ho G~C, Horbury T~S,
  Hu Q, Huang J, Klein K~G, Korreck K~E, Larson D~E, Livi R, Maruca B, Lavraud
  B, Louarn P, Maksimovic M, Martinovic M, McGinnis D, Pogorelov N~V,
  Richardson J~D, Skoug R~M, Steinberg J~T, Stevens M~L, Szabo A, Velli M,
  Whittlesey P~L, Wright K~H, Zank G~P, MacDowall R~J, McComas D~J, McNutt R~L,
  Pulupa M, Raouafi N~E and Schwadron N~A 2019 {\em Nature\/} {\bf 576} 228

\bibitem{kim.et07}
Kim Y~H, Moon Y~J, Park Y~D, Sakurai T, Chae J, Cho K~S and Bong S~C 2007 {\em
  Publications of the Astronomical Society of Japan\/} {\bf 59} 763

\bibitem{joshi.et17a}
Joshi N~C, Sterling A~C, Moore R~L, Magara T and Moon Y~J 2017 {\em
  Astrophysical Journal\/} {\bf 845} 26

\bibitem{sterling.et18}
Sterling A~C, Moore R~L and Panesar N~K 2018 {\em Astrophysical Journal\/} {\bf
  864} 68

\bibitem{raouafi.et14}
Raouafi N~E and Stenborg G 2014 {\em Astrophysical Journal\/} {\bf 787} 118

\bibitem{panesar.et18b}
Panesar N~K, Sterling A~C, Moore R~L, Tiwari S~K, De~Pontieu B and Norton A~A
  2018 {\em Astrophysical Journal\/} {\bf 868L} 27

\bibitem{panesar.et16b}
Panesar N~K, Sterling A~C and Moore R~L 2016 {\em Astrophysical Journal\/} {\bf
  822L} 7

\bibitem{panesar.et19}
Panesar N~K, Sterling A~C, Moore R~L, Winebarger A~R, Tiwari S~K, Savage S~L,
  Golub L~E, Rachmeler L~A, Kobayashi K, Brooks D~H, Cirtain J~W, De~Pontieu B,
  McKenzie D~E, Morton R~J, Peter H, Testa P, Walsh R~W and Warren H~P 2019
  {\em Astrophysical Journal Letters\/} {\bf 887} 8

\bibitem{beckers68}
Beckers J~M 1968 {\em Solar Physics\/} {\bf 3} 367

\bibitem{sterling00}
Sterling A~C 2000 {\em Solar Physics\/} {\bf 196} 79

\bibitem{depontieu.et07a}
{De Pontieu} B, {McIntosh} S, {Hansteen} V~H, {Carlsson} M, {Schrijver} C~J,
  {Tarbell} T~D, {Title} A~M, {Shine} R~A, {Suematsu} Y, {Tsuneta} S,
  {Katsukawa} Y, {Ichimoto} K, {Shimizu} T and {Nagata} S 2007 {\em
  Publications of the Astronomical Society of Japan\/} {\bf 59} 655

\bibitem{tsiropoula.et12}
Tsiropoula G, Tziotziou K, Kontogiannis I, Madjarska M~S, Doyle J~G and
  Suematsu Y 2012 {\em Space Science Reviews\/} {\bf 169} 181

\bibitem{samanta.et19}
Samanta T, Tian H, Yurchyshyn V, Peter H, Cao W, Sterling A, Erd{'e}lyi R, Ahn
  K, Feng S, Utz D, Banerjee D and Chen Y 2019 {\em Science\/} {\bf 366} 890

\bibitem{iijima.et17}
Iijima H and Yokoyama T 2017 {\em Astrophysical Journal\/} {\bf 848} 38

\bibitem{martinezsykora.et17}
Mart{\'i}nez-Sykora J, De~Pontieu B, Hansteen V~H, Rouppe van~der Voort L,
  Carlsson M and Pereira T~M~D 2017 {\em Science\/} {\bf 356} 1269

\bibitem{sterling.et16a}
Sterling A~C and Moore R~L 2016 {\em Astrophysical Journal\/} {\bf 828} L9

\bibitem{sterling.et20a}
Sterling A~C, Moore R~L, Samanta T and Yurchyshyn V 2020 {\em Astrophysical
  Journal Letters\/} {\bf 893} 45

\bibitem{pasachoff.et68}
Pasachoff J~M, Noyes R~W and Beckers J~M 1968 {\em Solar Physics\/} {\bf 5} 131

\bibitem{depontieu.et12}
{De Pontieu} B, {Carlsson} M, {Rouppe van der Voort} L~H~M, {Rutten} R~J,
  {Hansteen} V~H and {Watanabe} H 2012 {\em Astrophysical Journal\/} {\bf 752L}
  12

\bibitem{samanta.et15}
Samanta T, Pant V and Banerjee D 2015 {\em Astrophysical Journal\/} {\bf 815}
  L16

\bibitem{pant.et15}
Pant V, Dolla L, Mazumder R, Banerjee D, Krishna~Prasad S and Panditi V 2015
  {\em Astrophysical Journal\/} {\bf 807} 71

\bibitem{depontieu.et07c}
De~Pontieu B, McIntosh S~W, Carlsson M, Hansteen V~H, Tarbell T~D, Schrijver
  C~J, Title A~M, Shine R~A, Tsuneta S, Katsukawa Y, Ichimoto K, Suematsu Y,
  Shimizu T and Nagata S 2007 {\em Science\/} {\bf 218} 1574

\bibitem{jess.et09}
Jess D~B, Mathioudakis M, Erd{\' e}lyi R, Crockett P~J, Keenan F~P and
  Christian D~J 2009 {\em Science\/} {\bf 323} 5921

\bibitem{kuridze.et12}
Kuridze D, Morton R~J, Erd{\' e}lyi R, Dorrian G~D, Mathioudakis M, Jess D~B
  and Keenan F~P 2012 {\em Astrophysical Journal\/} {\bf 750} 51

\bibitem{kuridze.et13}
Kuridze D, Verth G, Mathioudakis M, Erd{\' e}lyi R, Jess D~B, Morton R~J,
  Christian D~J and Keenan F~P 2013 {\em Astrophysical Journal\/} {\bf 779} 82

\bibitem{morton.et13}
Morton R~J, Verth G, Fedun V, Shelyag S and Erd{\' e}lyi R 2013 {\em The
  Astrophysical Journal\/} {\bf 768} 17

\bibitem{morton14}
Morton R~J 2014 {\em Astronomy \& Astrophysics\/} {\bf 566} 90

\bibitem{zaqarashvili.et09}
Zaqarashvili T~V and Erdl{\'e}yi R 2009 {\em Solar Physics\/} {\bf 149} 355

\bibitem{tenerani.et20}
Tenerani A, Velli M, Matteini L, R{\'e}ville V, Shi C, Bale S~D, Kasper J~C,
  Bonnell J~W, Case A~W, de~Wit T~D, Goetz K, Harvey P~R, Klein K~G, Korreck K,
  Larson D, Livi R, MacDowall R~J, Malaspina D~M, Pulupa M and Stevens
  Michael~Whittlesey P 2020 {\em Astrophysical Journal Supplement Series\/}
  {\bf 246} 32

\bibitem{lionello.et16}
Lionello R, T{\" o}r{\" o}k T, Titov V~S, Leake J~E, Miki{\' c} Z, Linker J~A
  and Linton M~G 2016 {\em Astrophysical Journal Letters\/} {\bf 831L} 2L

\bibitem{roberts.et18}
Roberts M~A, Uritsky V~M, DeVore C~R and Karpen J~T 2018 {\em Astrophysical
  Journal\/} {\bf 866} 14

\bibitem{mozer.et20}
Mozer F~S, Agapitov O~V, Bale S~D, Bonnell J~W, Case T, Chaston C~C, Curtis
  D~W, Dudok~de Wit T, Goetz K, Goodrich K~A, Harvey P~R, Kasper J~C, Korreck
  K~E, Krasnoselskikh V, Larson D~E, Livi R, MacDowall R~J, Malaspina D, Pulupa
  M, Stevens M, Whittlesey P~L and Wygant J~R 2020 {\em Astrophysical Journal
  Supplement Series\/} {\bf 246} 68

\end{thebibliography}




\end{document}